\documentclass[twocolumn,showpacs,preprintnumbers,amsmath,amssymb,floatfix]{revtex4}

\usepackage{graphicx}
\usepackage{dcolumn}
\usepackage{bm}

\newcommand{\pom}{\tt I\! P}
\newcommand{\beq}{\begin{equation}}
\newcommand{\eeq}{\end{equation}}

\begin{document}

\title{Charm and bottom production in inclusive double Pomeron exchange in heavy ion collisions at the LHC}


\author{M. B. Gay Ducati, M. M. Machado and M. V. T. Machado}

\affiliation{High Energy Physics Phenomenology Group, GFPAE,  IF-UFRGS \\
Caixa Postal 15051, CEP 91501-970, Porto Alegre, RS, Brazil}

\begin{abstract}

The inclusive double Pomeron exchange cross section for heavy quark pair production is calculated for nucleus-nucleus collisions at the LHC.  The present estimate is based on hard diffractive factorization, corrected by absorptive corrections and nuclear effects. The theoretical uncertainties for nuclear collisions are investigated and a comparison to other approaches is presented. The production channels giving a similar final state configuration are discussed as well.
\end{abstract}

\pacs{13.60.Hb, 12.38.Bx, 12.40.Nn, 13.85.Ni, 14.40.Gx}

\maketitle

\section{Introduction}

At the LHC, heavy quark pairs are produced in large quantities and they are important both for perturbative QCD physics study and for understanding the background to other processes. For instance, heavy flavored hadrons may also produce high momentum leptons, and therefore they always constitute a potential background to new physics. The theory of hadronic production of heavy quarks is in reasonably good shape. The inclusive production is known beyond the   next-to-leading level order in QCD and the results are considered well established. Recently, the clean topologies of exclusive particle production in electromagnetic interactions in context of hadron-hadron and nucleus-nucleus collisions mediated by colorless exchanges such the QCD Pomeron have attracted an increasing interest \cite{upcs}. The cross sections for these processes are smaller than the correspondent inclusive production channels, which it is compensated by a more favorable signal/background relation. Experimentally, exclusive events are identified by  large rapidity gaps on both sides of the produced central system and the survival of both initial state particles scattered at very forward angles with respect to the beam.

Our goal in this work is to estimate the inclusive double Pomeron exchange cross section for heavy quarks production in heavy ion collisions at the LHC.  We focus on the coherent diffractive  production of heavy quarks in relativistic nucleus-nucleus collisions, i.e. the process $A+A\rightarrow A+[LRG]+[Q\bar{Q}+X]+[LRG]+A$, where $[LRG]$ stands for Large Rapidity Gap. Notice that the current study is complementary to our previous investigations on single diffractive production of charm/bottom presented in Ref. \cite{GMM}. It is also possible to occur incoherent diffractive scattering, where one or both nuclei will be excited, that subsequently decays into a system of colorless protons, neutrons and nuclei debris. Such processes are of experimental interest at the LHC, for instance at ALICE experiment \cite{ALICE}. Here, we start by the hard diffractive factorization, where the diffractive cross section is the convolution of the diffractive parton distribution functions and the corresponding diffractive coefficient functions, in a similar way as for the inclusive case. To further correct this approach it is introduced the appropriate  absorptive (unitarity) effects which cause the suppression of any large rapidity gap process being  important for the reliability of predictions \cite{k3p}. It is also considered the nuclear effects for nucleus-nucleus collisions and the predictions compared to other approaches available in the literature.

This paper is organized as follows. In the next section it is summarized the main formulas considered to compute the inclusive double Pomeron exchange (DPE) cross section and diffractive ratios for the hadroproduction of charm and bottom. We also present the procedure to evaluate the nuclear coherent diffractive cross sections at the LHC energies. At this point it is introduced other models for the inclusive diffractive heavy quark production.  In the last section, the numerical results are shown for the inclusive and diffractive cross sections as a function of energy and give predictions for the corresponding diffractive ratios for $pp$ and $AA$ collisions. Discussion on the nuclear dependence of cross sections and corresponding suppression factors are  addressed. A comparison to competing channel, as the exclusive photo-nuclear reactions in $AA$ collisions and two-photon production, is performed.

%
%
\section{Heavy-quark production in inclusive DPE process}

Let us present the main formulas for the inclusive diffractive cross sections for the production of heavy quarks in proton-proton collisions at high energies. We rely on diffractive factorization formalism \cite{IS} and take into account both the absorption effects (multiple Pomeron exchange) and nuclear effects using Glauber based approach. The limitations and theoretical incompleteness of such a picture are well known. However, it is reasonable for a first exploratory study. In the inclusive case, the process is described for partons of two protons, interacting to produce a heavy quark pair, $p+p\rightarrow Q\bar{Q}+X$, with center of mass energy $\sqrt{s}$. At LHC energies, the gluon fusion channel dominates over the $q\bar{q}$ annihilation process and $qg$ scattering. The NLO cross section is obtained by convoluting the partonic cross section with the parton distribution function (PDF), $g(x,\mu_F)$, in the proton, where $\mu_F$ is the factorization scale. At any order, the partonic cross section may be expressed in terms of dimensionless scaling functions $f^{k,l}_{ij}$ that depend only on the variable $\rho$ \cite{13magno},
\begin{eqnarray}
\hat{\sigma}_{ij}(\hat{s},m^{2}_{Q},\mu^{2}_{F},\mu^{2}_{R})   & = & \frac{\alpha^{2}_{s}(\mu_{R})}{m^{2}_{Q}}\sum_{k=0}^{\infty}\left [ 4\pi\alpha_{s}(\mu_{R})\right ] ^{k} \nonumber \\
& \times &\sum^{a}_{l=0}f^{(k,l)}_{ij}(\rho)\ln^{l}\left ( \frac{\mu^{2}_{F}}{m^{2}_{Q}}\right )\,,
\label{equacao1}
\end{eqnarray}
where $\rho=\frac{\hat{s}}{4m^{2}_{Q}-s_0}$, $i,j=q,\bar{q},g$, specifying the types of the annihilating partons, $\hat{s}$ is the partonic center of mass, $m_{Q}$ is the heavy quark mass, $\mu_{R}$ is the renormalization scale ($s_0=1$ GeV$^2$). It is calculated as an expansion in powers of $\alpha_{s}$ with $k=0$ corresponding to the Born cross section at order ${\cal O}(\alpha^{2}_{s})$. The first correction, $k=1$, corresponds to the NLO cross section at ${\cal O}(\alpha^{3}_{s})$. To calculate the $f_{ij}$ in perturbation theory, both renormalisation and factorisation scale of mass singularities must be performed. The subtractions required are done at the mass scale $\mu$. The running of the coupling constant $\alpha_{s}$ is determined by the renormalization group. The total hadronic cross section for the heavy quark production is obtained by convoluting the total partonic cross section with the parton distribution functions of the initial hadrons \cite{nason}
\begin{eqnarray}\nonumber
\sigma_{pp}(s,m^{2}_{Q}) & = & \sum_{i,j}\int^{1}_{\tau}dx_{1}\int^{1}_{\frac{\tau}{x_{1}}}dx_{2}f^{p}_{i}(x_{1},\mu^{2}_{F})f^{p}_{j}(x_{2},\mu^{2}_{F}) \\  & \times  & \hat{\sigma}_{ij}(\hat{s},m^{2}_{Q},\mu^{2}_{F},\mu^{2}_{R}),
\label{Matrix}
\end{eqnarray}
with the sum $i,j$ over all massless partons. Here, $x_{1,2}$ are the hadron momentum fractions carried by the interacting partons, $f^{p}_{i(j)}$ is the parton distribution functions, evaluated at the factorization scale and assumed to be equal to the renormalization scale in our calculations. Here, the cross sections were calculated with the following mass and scale parameters: $\mu_{c}=2m_{c}$, $m_{c}=1.5$ GeV, $\mu_{b}=m_{b}=4.5$ GeV, based on the current phenomenology for heavy quark hadroproduction \cite{15magno}.

For diffractive processes, we rely on the hard diffractive factorization \cite{IS}, where
the Pomeron structure (quark and gluon content) is probed. In the case of single diffraction, a Pomeron is emitted by one of the colliding hadrons. That hadron is detected, at least in principle, in the final state and the remaining hadron scatters off the emitted Pomeron. A typical single diffractive reaction is given by $p+p\rightarrow p+Q\bar{Q}+X$, with the cross section assumed to factorise into the total Pomeron--hadron cross section and the  Pomeron  flux  factor  \cite{IS}, $f_{{\rm\pom}/i}(x^{(i)}_{\pom},|t_i|)$.  As usual, the Pomeron kinematical variable $x_{\pom}$ is defined as $x_{\pom}^{(i)}=s_{\pom}^{(j)}/s_{ij}$, where  $\sqrt{s_{\pom}^{(j)}}$ is the center-of-mass energy in the Pomeron--hadron $j$ system and $\sqrt{s_{ij}}=\sqrt{s}$ the center-of-mass energy in the hadron $i$--hadron $j$ system. The momentum transfer in the  hadron $i$ vertex is denoted by $t_i$. A similar approach can also be applied to double Pomeron exchange (DPE) process, where both colliding hadrons  can in  principle  be detected in the final  state.  Thus, a typical reaction would be  $p+p \rightarrow p+ Q\bar{Q} + X+p$, and DPE  events are characterized by two  quasi--elastic hadrons with  rapidity  gaps between them and the central heavy flavor products. The inclusive DPE cross section may then be written as,
\begin{eqnarray}
 \label{sddexp}
\frac{d\sigma(pp\rightarrow pp+ Q\bar{Q}+X)}
{dx^{(1)}_{\pom}dx^{(2)}_{\pom}d|t_1|d|t_2|}\!\! & = & \!\! f_{{\rm\pom}/p}(x^{(1)}_{\pom},|t_1|)\, f_{{\rm\pom}/p}(x^{(2)}_{\pom},|t_2|) \nonumber\\
\! & \times & \sum_{i,j=q,g}\!\sigma\left({\pom} + {\pom}\rightarrow  Q\bar{Q}  +  X\right),\nonumber \\
\end{eqnarray}
where the Pomeron-Pomeron cross section is given by,
\begin{eqnarray}
\label{ddxsect}
\sigma\left({\pom} + {\pom}\rightarrow  Q\bar{Q}  +  X\right) &=& \int\int dx_1 \,dx_2 \,\hat \sigma_{ij}(\hat{s} ,m_Q^2,\mu^2) \nonumber \\
&\times & f_{i/\pom}\left(\beta_1,\mu^2\right) f_{j/\pom}\left(\beta_2,\mu^2\right),\,\,\,\, \,\,\,\,
\end{eqnarray}
where $f_{i/\pom}\left(\beta,\mu^2\right)$ are the diffractive parton (quark, gluon) distribution functions (DPDFs) evaluated for parton   momentum fraction $\beta_{a}=x_{a}/x_{\pom}^{a}$ ($a=1,2$) and evolution scale $\mu^2$.

We further correct Eq. (\ref{sddexp})  by considering the suppression of the hard diffractive cross section by multiple-Pomeron scattering effects (absorptive corrections). This is taken into account through a gap survival probability, $S_{\mathrm{gap}}^2$, which can be described in terms of screening or absorptive corrections \cite{Bj}.  There are intense theoretical investigations on this subject in the last years and we quote Ref. \cite{Prygarin} for a discussion on several theoretical estimations for the gap survival probabilities. As a baseline value, we follow  Ref. \cite{KKMR}. For the present purpose, we consider $S_{\mathrm{gap}}^2=0.032\,(0.031)$ at $\sqrt{s}=5.5\, (6.3)$ TeV in nucleon-nucleon collisions, which is obtained using  a parametric interpolation formula for the KMR survival probability factor \cite{KKMR} in the form $ S_{\mathrm{gap}}^2  =a/[b+\ln (\sqrt{s/s_0})]$ with $a = 0.126$, $b=-4.688$ and $s_0=1$ GeV$^2$. This formula interpolates between survival probabilities for central diffraction (CD) in  proton-proton collisions of $4.5 \, \%$ at Tevatron and $2.6 \, \%$ at the LHC.

For the heavy quarks production in nucleus-nucleus collisions, two processes can occur: (a) coherent diffractive scattering, $A+A\rightarrow A+Q\bar{Q}+X+A$, and (b) incoherent diffractive scattering, $A+A\rightarrow A^*+Q\bar{Q}+X+A^{(*)}$.   In the first case,  both nuclei emit Pomerons and partons from them interact with each other. Thus, both nuclei remain intact at the final state. For the incoherent diffractive scattering, one or both nuclei can be excited, represented by $A^*$, that subsequently decay into a system of colorless protons, neutrons and nuclei debris. Here, we focus on the coherent case. In order to calculate the $AA$ cross section the procedure presented originally in Ref. \cite{Pajares} (see Refs.  \cite{Pajares2} for further applications to soft processes), the so-called {\it criterion C}, will be used. The central diffraction (coherent) cross section for $AB$ collisions is given by:
\begin{eqnarray}
\sigma^{\mathrm{CD}}_{AB}  =  \sigma^{in}_{AB}\left(\sigma^{in}_{pp}\right)-\sigma^ {in}_{AB}\left(\sigma^{in}_{pp}-\sigma^{\mathrm{CD}}_{pp}\right),
\label{sigsdaa}
\end{eqnarray}
where $\sigma^{in}_{AB}$ is the inelastic $AB$ cross section considered as a function of the nucleon-nucleon total cross section $\sigma$ ($\sigma^{in}_{pp}$ and $\sigma^{\mathrm{CD}}_{pp}$ are the inelastic and CD cross sections in proton-proton case, respectively).  The application of the formalism to $AA$ collisions turns out to be difficult due to the absence of an explicit expression for $\sigma^{in}_{AA}(\sigma)$. In Ref. \cite{Pajares,Pajares2} the authors considered optical approximation in which $\sigma^{in}_{AA}$ is given by corresponding formula for $pA$ with $A\rightarrow AB$ and an effective profile function for two colliding nuclei at impact parameter $b$, $T_{AB}=\int d^2\bar{b}\,T_A(\bar{b})\,T_B(b-\bar{b})$. The final expression for CD  cross section in $AA$ collisions is given by
\begin{eqnarray}
\sigma^{\mathrm{DPE}}_{AA} =  A^2\int d^2b \,T_{AA}(b)\exp\left[-A^2\,\sigma^{in}_{pp}\,T_{AA}(b)\right]\sigma^{\mathrm{DPE}}_{pp},\label{sdxs}
\label{sdxs1}
\end{eqnarray}
where we consider Woods-Saxon nuclear densities and the inelastic cross section at the LHC is taken to be  $\sigma^{in}_{pp}\,(\sqrt{s}=6\,\mathrm{TeV})=73$ mb \cite{Pajares,Pajares2}.

To analyze the model dependence of the cross section, we consider another approach to inclusive diffractive production of heavy quarks. In order to do so, the  Bialas-Landshoff (BL) approach \cite{Land-Nacht,Bial-Land} for the process $p+p\rightarrow p+Q\bar{Q}+p$ is taken into account.  The calculation that follows concerns  central inclusive process, where the QCD radiation accompanying the produced object is allowed. Thus, we did not include a Sudakov survival factor $T(\kappa,\mu)$ \cite{Khoze} which is needed for exclusive central processes. The cross-section is given by \cite{Bial-Szer}:
\begin{equation}
\sigma_{\pom \pom}(\mathrm{BL})=\frac{1}{2s\left(  2\pi\right)  ^{8}}\int\overline{|M_{fi}|^{2}%
}\left[  F\left(  t_{1}\right)  F\left(  t_{2}\right)  \right]  ^{2}dPH,
\label{cro-sec-ogolny}%
\end{equation}
where $F\left(  t\right)  $ is the nucleon form-factor approximated by $F\left(  t\right)  =\exp\left(b\, t\right)$, with slope parameter $b=$ $2$ GeV$^{-2}$. The differential phase-space factor $dPH$ has the form,
\begin{eqnarray}
dPH  &  = & d^{4}k_{1}\delta\left(  k_{1}^{2}\right)  d^{4}k_{2}\,\delta\left(
k_{2}^{2}\right)  d^{4}r_{1}\,\delta\left(  r_{1}^{2}-m_Q^2\right) \nonumber \\
& \times & d^{4}r_{2}\,
\delta\left(  r_{2}^{2}-m_Q^2\right) \Theta\left(  k_{1}^{0}\right)  \Theta\left(  k_{2}^{0}\right)
\Theta\left(  r_{1}^{0}\right)  \Theta\left(  r_{2}^{0}\right) \nonumber \\
&  \times & \delta
^{(4)}\left(  p_{1}+p_{2}-k_{1}-k_{2}-r_{1}-r_{2}\right) ,
\end{eqnarray}
where $m_Q$ is the mass of the produced quarks. Following
\cite{Bial-Szer}, the use of Sudakov parameterization for momenta is given by
\begin{eqnarray}
Q  &  = &\frac{x}{s}p_{1}+\frac{y}{s}p_{2}+v,\hspace{0.5cm}k_{1}    = x_{1}p_{1}+\frac{y_{1}}{s}p_{2}+v_{1},\nonumber\\
k_{2}  &  = & \frac{x_{2}}{s}p_{1}+y_{2}p_{2}+v_{2},\hspace{0.5cm} r_{2}    =  x_{Q}p_{1}+y_{Q}p_{2}+v_{Q}, \nonumber
\end{eqnarray}
where $v,$ $v_{1},$ $v_{2},$ $v_{Q}$ are two-dimensional four-vectors describing the transverse components of the momenta. The momenta for the incoming (outgoing) protons are $p_1,\,p_2$ ($k_1,\,k_2$) and the momentum for the produced quark (antiquark) is $r_2$ ($r_1$), whereas the momentum for one of the exchanged gluons is $Q$. The square of the invariant matrix element averaged over initial spins and summed over final spins is given by \cite{Bial-Szer},
\begin{eqnarray}
\overline{\left|  M_{fi}\right|  ^{2}} & = & \frac{x_1y_2\,H}{\left(sx_{Q}y_{Q}\right)
^{2}\left(  \delta_{1}\delta_{2}\right)  ^{1+2\epsilon}\delta_{1}^{2\alpha
^{\prime}t_{1}}\delta_{2}^{2\alpha^{\prime}t_{2}}}\nonumber \\
&\times & \left(1-\frac{4\,m_Q^2}{s\delta_1\delta_2}\right)\,\exp\left[  2\beta\left(
t_{1}+t_{2}\right)  \right]  .
\end{eqnarray}

In the expression above, $\delta_{1}=1-x_1$,  $\delta_{2}=1-y_2$, $t_{1}=-\vec{v}_{1}^{2}$ and $t_{2}=-\vec{v}_{2}^{2}$. The factor $\exp\left[ 2\beta\left(  t_{1}+t_{2}\right)  \right]$ takes into account the effect of the momentum transfer dependence of the non-perturbative gluon propagator with $\beta=1$ GeV$^{-2}$. The overall normalization can be expressed as,
\begin{eqnarray}
H = S_{\mathrm{gap}}^2\times2s\,\left[ \frac{4\pi m_Q\,(G^2D_0)^3\mu^4}{9\,(2\pi)^2}  \right]^2\,\left(\frac{\alpha_s}{\alpha_{0}}\right)^2,
\end{eqnarray}
where $\alpha_s$ is the perturbative coupling constant (it depends on the hard scale) and $\alpha_0$ (supposed to be independent of the hard scale) is the unknown nonperturbative coupling constant. In the numerical calculation, we use the parameters \cite{Bial-Szer} $\epsilon=0.08,$ $\alpha^{\prime}=0.25$ GeV$^{-2}$, $\mu=1.1$ GeV and $G^{2}D_{0}=30$ GeV$^{-1}\mu^{-1}$. The Regge Pomeron trajectory is then $\alpha_{\pom}(t)= 1+\epsilon+\alpha^{\prime} t$. It is taken $k_{\mathrm{min}} = 0$ for the minimum value for the transverse momentum of the quark. For the strong coupling constant, we use $\alpha_s=0.2 \,(0.17)$ for charm (bottom). An indirect determination of the unknown parameter $\alpha_0$ has been found in Ref. \cite{Adam} using experimental data for central inclusive dijet production cross section at Tevatron. Namely,  it has been found the constraint $S_{\mathrm{gap}}^2\,(\sqrt{s}=2\,\mathrm{TeV})/\alpha_0^2 = 0.6$, where $S_{\mathrm{gap}}^2$ is the gap survival probability factor (absorption factor). Considering the KMR \cite{KKMR} value $S_{\mathrm{gap}}^2=0.045$ for CD processes at Tevatron energy, one obtains $\alpha_0^2=0.075$. For the nuclear version of the cross section we use the same procedure described before.

\begin{table}[t]
\centering
\renewcommand{\arraystretch}{1.5}
\begin{tabular}{c c c c}
\hline
	$Q\bar{Q}$       &   $\sigma_{\mathrm{inc}}$ $[\mu b]$    &	$\sigma_{\mathrm{DPE}}$ $[\mu b]$ 	&	$R_{\mathrm{DPE}}$ [\%] \\\hline
$c\bar{c}$	&	7811           &  13.6--0.53      	&	0.17--7$\times10^{-3}$      \\
$b\bar{b}$	&	393      	    &	0.053--0.027         	&	0.01--0.007     \\
\hline
\end{tabular}
\caption{The inclusive and DPE (corrected by absorption effects) cross sections in $pp$ collisions at the LHC. For the inclusive diffractive cross section the first value corresponds to the Ingelman-Schlein approach and the second one the Bialas-Landshoff approach. The corresponding diffractive ratios, $R_{\mathrm{DPE}}$, are also presented.}
\label{tabelahadronica}
\end{table}

In what follows, we present the numerical results using procedures referred above and compare them to distinct theoretical approaches in literature. Competing channels producing the same final state configuration (two large rapidity gaps and central produced system) are also  discussed.

\subsection{Results and discussion}

First, we present the results for the inclusive and diffractive heavy quarks cross  sections for hadronic collisions at LHC. The calculations for the inclusive and diffractive cross sections as well as the diffractive ratios to heavy quark production in proton-proton collisions are showed at Tab. (\ref{tabelahadronica}). For the inclusive diffractive cross section the first value corresponds to the partonic picture of Pomeron,  Eq. (\ref{sddexp}), and the second one to the Bialas-Landshoff approach, Eq. (\ref{cro-sec-ogolny}). We assume the value $S_{\mathrm{gap}}^2=0.026$ for the absorption corrections at energy of 14  TeV. The partonic PDFs and scales are mentioned in previous section. For the diffractive gluon PDF, we take the experimental (H1 collaboration) FIT A \cite{H1diff}. The main theoretical uncertainty in the diffractive ratio is the survival probability factor, whereas uncertainties associated to factorization/renormalization scale, parton PDFs and quark mass are minimized taking a ratio. The present results are consistent with a previous estimate performed in Ref. \cite{magno}, where a value $S_{\mathrm{gap}}^2=0.04$ was considered and cross sections were computed at LO accuracy.

\begin{table*}[t]
\centering
\renewcommand{\arraystretch}{1.5}
\begin{tabular}{c c c c c c }
\hline
                        &   CaCa  [$c\bar{c}$]  &      PbPb [$c\bar{c}$] &    CaCa [$b\bar{b}$] &   PbPb [$b\bar{b}$]  \\\hline
 $\sigma_{AA}^{\mathrm{DPE}}$ [$\mu$b]  & 22.8--2.8  & 31.1--4.2   &  0.25--0.14              &  0.32--0.2  \\
$R_{\mathrm{coh}}^{\mathrm{DPE}} [\%]$          &    $3-0.4\times 10^{-4}$              &   $2-0.2\times 10^{-4}$          & $8-4\times 10^{-5}$                 &   $4-3\times 10^{-6}$  	      \\ \hline
\end{tabular}
\caption{The diffractive coherent cross section for CaCa and PbPb collisions. The first value corresponds to the partonic Pomeron approach and the second one to the Bialas-Landshoff approach. Diffractive ratios are also presented.}
\label{tabelaCoerente}
\end{table*}

It is important to discuss the uncertainties in our estimates presented in Tables I and II. In the partonic Pomeron model, the main theoretical uncertainties come from the factorization/renormalization scale and from the diffractive PDFs. The scale dependence is stronger for charm case (the error band reaches a factor around 2 \cite{15magno}) and very stable for bottom. We have checked that the cross sections are insensitive to a distinct choice on the diffractive gluon  PDF (for instance, H1 Collaboration, FIT B). The main theoretical uncertainty in the Bialas-Landshoff model is the nonperturbative coupling $\alpha_0$. We have used the adjusted value from \cite{Adam}, extracted from experimental data for central inclusive dijet cross section at $\sqrt{s}=1.8$ TeV and we believe that it is stable for extrapolations to high energies. A considerable concerning in our calculation is the assumption of the gap survival probability in $AA$ collisions being the the same as for $pp$ collisions and in addition to be final-state independent. The last point is better understood than the first one. Considering for simplicity the usual eikonal (one channel) model \cite{Maor}, it is known that the gap survival probability for DPE processes is given by $S_{\mathrm{gap}}^2=a[2\nu(s)]^{-a}\gamma (a,2\nu(s))$, where $\gamma(a,2\nu)$ denotes the incomplete Euler gamma function. The variable $a$ depends on energy, $s$, and on final state mass, $M$. Namely, $a(s,M)=2R^2(s)/R^2(\frac{s}{M^2})$, with $R = 4[R_0^2+\alpha^{\prime}\ln (s/s_0)]$. In addition, $\nu(s)=\frac{\sigma_0}{2\pi R^2(s)}(s/s_0)^{\epsilon}$. It is clear that $a(s,M)$ has a rather weak dependence on $M^2$ as it is proportional to $\alpha^{\prime}\ln (M^2)$ (with $\alpha^{\prime}$ small) over a relatively narrow domain. Therefore, it can be factored out of $M^2$ integration. For a numerical example of the small sensitivity of final-state configuration, in \cite{Maor} the gap survival probability is computed for hard diffractive dijets and for meson $\chi_c(3415)$ and the deviation is shown to be of order 5\% at 14 TeV. The question about the gap survival in heavy-ion collisions is an open question in literature. It is naively expected that it depends on impact parameter and  will be different in $pp$ and $AA$ collisions. For the $pp$ case, it has been shown in Ref. \cite{SW} that the uncertainties on the numerical predictions for $S_{\mathrm{gap}}^2$ come from the transverse spatial distribution of gluons entering in $P_{hard}(b)$. This quantity is the probability for two gluons to collide at same transverse point as a function of the $pp$ impact parameter, given by the convolution of the transverse spatial distributions of the gluons in the colliding protons. An additional source of uncertainty arises from the modeling of the $pp$ elastic amplitude, $\Gamma (b)$. It was also shown in \cite{SW} that moderate extra suppression results from fluctuations of the partonic configurations of the colliding protons and at LHC energies absorptive interactions of hard spectator partons associated with the process of gluon fusion producing a massive final state reach to the black-disc regime and it would generates sizable additional suppression. A careful analysis along these line was not still done for $AA$ collisions.

Let us discuss the model dependence of the cross sections. The partonic approach for Pomeron produces a larger cross section compared to Bialas-Landshoff model. This comes mostly from the distinct energy dependence in the phenomenological models. The first one considers a semihard Pomeron intercept fitted from HERA diffractive DIS data,  where $\epsilon \simeq 0.12$. On the other hand, the soft Pomeron models consider the conservative soft Pomeron intercept, $\epsilon = 0.08$. The deviation less evident for bottom may come from a distinct $m_Q$ dependence on each approach.  The present calculation can be compared to the exclusive diffractive (central diffraction, CD) channel, $p+p\rightarrow p+Q\bar{Q}+p$. Very recently, in Refs. \cite{Antony} it has been computed considering the KKMR procedure to compute the CD processes for open charm and bottom production.  Notice that the Ingelman-Schlein approach (IS) gives larger cross sections so far as it refers to the inclusive central diffraction, $p+p\rightarrow p+Q\bar{Q}X+p$, where the remnants in the final state of the Pomeron-Pomeron sub-system ($X$) would share the energy and momentum continuously, and therefore no distinct gap would emerge.

As a cross-check for Tevatron at $\sqrt{s}=1.96$ TeV we got  1.08--0.37 $\mu$b for charm and 6--16 nb for bottom. These values can be compared to the exclusive heavy quark production recently computed using the KKMR procedure \cite{Antony}. In that case, the imaginary part of the amplitude of the exclusive diffractive quark pair production, $p+p\rightarrow p+Q\bar{Q}+p$, is given by \cite{Antony},
\begin{eqnarray}
& & {\cal M}_{\lambda_Q\lambda_{\bar{Q}}} =  \frac{s\,\pi^2\delta_{c_1c_2}}{2(N_c^2-1)}\int d^2\kappa_0 \frac{V_{\lambda_Q\lambda_{\bar{Q}}}^{c_1c_2}(\kappa_1,\kappa_2,\ell_1,\ell_2)}{\kappa_0^2\,\kappa_1^2\,\kappa_2^2}\nonumber\\
& & \times \, {\cal F}_g(x_1,x_1^{\prime},\kappa_0^2,\kappa_1^2,t_1)\,{\cal F}_g(x_2,x_2^{\prime},\kappa_0^2,\kappa_2^2,t_2),
\end{eqnarray}
where $\lambda_Q$ and $\lambda_{\bar{Q}}$ are helicities of heavy $Q$ and $\bar{Q}$, respectively. Above, the quantities ${\cal F}_g$ are the off-diagonal unintegrated gluon distributions (UGDs) in each nucleon, $\kappa_{0,1,2}$ ($x_1,x_2$) are the gluon transverse momenta (longitudinal momentum fractions of active gluons) in the KKMR approach and $\ell_{1,2}$ are the quark (antiquark) momenta. The vertex factor  $V_{\lambda_Q\lambda_{\bar{Q}}}^{c_1c_2}(\kappa_1,\kappa_2,\ell_1,\ell_2)$ is the production amplitude of a pair of massive quarks with helicities $\lambda_{Q,\bar{Q}}$ and momenta $\ell_{1,2}$. There is a lot of uncertainties concerning the off-diagonal UGDs even in the nucleon case.  After some exercise, we are able to estimate that in that work the cross section is $\approx$ 2.97--1.1 $\mu$b for charm case (at Tevatron) and $\approx 1.3$ nb for bottom at the LHC.

Let us move now to the computation of nucleus-nucleus cross sections. The results are shown in Tab. \ref{tabelaCoerente} for diffractive coherent  collisions. The corresponding diffractive ratio is obtained from
\begin{eqnarray}
R_{\mathrm{coh}}^{\mathrm{DPE}}(\sqrt{s})=\frac{\sigma_{AA}^{\mathrm{DPE}}(\sqrt{s})}{\sigma_{AA}^{\mathrm{inc}}(\sqrt{s})}\times \,100\%,
\label{equacaoIC}
\end{eqnarray}
where $\sigma_{AA}^{\mathrm{inc}}$ is given by the hadroproduction cross section in Eq. (\ref{Matrix}) and using minimum bias cross section for $AA$ collisions. The diffractive DPE cross section in nucleus-nucleus collisions is obtained from Eq. (\ref{sdxs1}).

In Tab. (\ref{tabelaCoerente}), the coherent cross section is presented for calcium and lead nuclei. The general trend on their magnitudes follows the main features occurring at the proton-proton case. Namely, larger deviation between IS and  BL models for the charm case, whereas the difference  diminishes for bottom. The diffractive ratios are quite small compared to the proton case. The present calculation is still somewhat consistent with earlier calculation presented in Ref. \cite{Agababyan} (absorption is not take into account in that work), which predicts $\sigma^{\mathrm{DPE}}=0.02-60$ $\mu$b for charm and $\sigma^{\mathrm{DPE}}= 0.003-1.4$ $\mu$b for bottom in PbPb collisions. The nuclear effect is proportional to $A^{1/3}$, which enhances by one order of magnitude the cross section compared to the nucleon case. The incoherent case is not addressed here, but a procedure to its estimation  has been presented in our previous study on single diffractive quark pair production \cite{GMM}. As the nuclear shadowing correction is an important theoretical uncertainty in the current work, it is timely to compare the predictions with different models. One possible alternate approach is the formalism considered by Muller and Schramm \cite{Schramm} for the meson (and Higgs) production in double Pomeron exchange. In that case, the cross section for particle production via two Pomerons exchange is written as
\begin{eqnarray}
\sigma_{AA}^{\pom\pom}= \int \int dx_1\,dx_2\,f_{\pom}(x_1)\,f_{\pom}(x_2)\,\sigma_{\pom\pom}(s_{\pom\pom}),
\end{eqnarray}
where $f_{\pom}(x)$ is the distribution function that describes the probability of finding a Pomeron in the nucleus with energy fraction $x$ and $\sigma_{\pom\pom}(s_{\pom\pom})$ is the subprocess cross section with squared energy, $s_{\pom\pom}$. It has been shown in \cite{Schramm} that function $f_{\pom}$, integrated over momentum transfer $t$, can be written as
\begin{eqnarray}
f_{\pom}(x)= \left(\frac{3\,A\beta_0\,Q_0}{2\pi}\right)^2\,\frac{1}{x}\,\left(\frac{s^{\prime}}{m_p^2}\right)^{2\epsilon}\exp\left(\frac{x^2M_A^2}{Q_0^2}\right),
\end{eqnarray}
with the soft Pomeron parameters $\epsilon =0.085$ and $\beta_0=1.8$ GeV$^{-1}$. Assuming a Gaussian expression for the nuclear form factor in impact parameter space, one has $Q_0=60$ MeV. In addition, $m_p$ and $M_A$ are the proton and nucleus masses, respectively. The differential cross section in impact parameter space is then given by,
\begin{eqnarray}
\frac{d^2\sigma_{AA}^{{\pom\pom}\rightarrow X}}{d^2b} & = &\left(\frac{3A\beta_0Q_0^2}{2\pi^2}\right)^4\int \int \frac{dx_1}{x_1}\frac{dx_2}{x_2}\,(2\pi\, b^2\,Q_0^4) \nonumber \\
& \times & \exp \left[ -\frac{M_A^2}{Q_0^2}\left(x_1^2+x_2^2\right)  \right] \exp \left(-\frac{b^2Q_0^2}{2}\right) \nonumber \\
& \times & \left(\frac{x_1x_2s^2}{m_p^4}\right)^{2\epsilon}\,\sigma_{AA}^{{\pom\pom}\rightarrow X}\left(x_1x_2s\right).
\end{eqnarray}
To compute the cross section for the subprocess $\pom\pom \rightarrow X$, authors of Ref.  \cite{Schramm} rely on the Pomeron model of Donnachie-Landshoff. Namely, it is assumed that the Pomeron couples to the quarks like an isoscalar photon. This means that the cross section of such a subprocess can be obtained from suitable modifications on the cross section for $\gamma\gamma\rightarrow X$. Another aspect to be considered is that the Pomeron-quark-quark vertex is not point-like, and when either or both of the two quark legs in this vertex goes far off shell, the coupling is known to decrease. So the quark-Pomeron coupling $\beta_0$ must be replaced by $\bar{\beta_0}(Q^2) = \beta_0\,\mu_0^2/(\mu_0^2+Q^2)$, where $\mu_0^2 = 1.2$ GeV$^2$ is a mass scale characteristic of the Pomeron; in our case of heavy quark production we have set $Q= m_Q$. Therefore, the process $\pom\pom \rightarrow X$ is totally similar to the one initiated by photons unless from an appropriate change of factors. The  cross section $\sigma ({\pom\pom}\rightarrow Q\bar{Q})$ is obtained by considering the cross section $\sigma (\gamma \gamma\rightarrow Q\bar{Q})$ by changing the fine-structure constant squared, $\alpha^2$, by $9\,\bar{\beta_0}^4/(16\pi^2)$. We consider the direct QED contribution to the process $\gamma\gamma\rightarrow Q\bar{Q}$, which should be dominant contribution  in the present purpose. The corresponding DPE cross section is given by
\begin{eqnarray}
\sigma ({\pom \pom} \rightarrow Q\bar{Q}) & = & \sigma_0\left[ \left( 1+\frac{4m_Q^2}{\hat{s}} -\frac{8m_Q^4}{\hat{s}^2}\right)\ln \frac{1+\omega}{1-\omega}\right. \nonumber \\
& - & \left. \omega \left( 1+\frac{4m_Q^2}{\hat{s}} \right) \right],
\end{eqnarray}
where $\sigma_0=27e_Q^2\bar{\beta}_0^4/(4\pi )$ and $\omega=\sqrt{1-(4m_Q^2/\hat{s})}$.
Integrating over impact parameter we found the following cross sections: 0.12 mb for charm and 0.8 nb for bottom in PbPb collisions at the LHC. This is consistent with the computed cross sections for DPE production of $\eta$-mesons in Ref. \cite{Schramm_etas}, which gives 0.76 mb and 0.81 nb for $\eta_c$ and $\eta_b$, respectively. After correcting  them by a gap survival probability one obtains $\sigma_{AA}^{\pom\pom}(c\bar{c})= 3.8$ $\mu$b and  $\sigma_{AA}^{\pom\pom}(b\bar{b})= 0.02$ nb. The strong suppression for bottom case comes from the model for the Pomeron-quark coupling, producing a suppression proportional to $(\mu_0^2+m_c^2)^4/(\mu_0^2+m_b^2)^4\simeq 7\times 10^{-4}$ plus the suppression due to distinct quark charge $e_b^4/e_c^4=1/16$. For the charm case, the Muller and Schramm model produces a similar result as for the Bialas-Landshoff model, whereas produces a smaller cross section for bottom case. 

\begin{table}[t]
\begin{tabular}{c c c c c }
\hline
 Heavy Quark  & $\pom\pom$ (IS) & $\pom\pom$ (BL) & $\gamma\gamma$ & $\gamma \pom$ \\
\hline
 $c\bar{c}$ & 32.5 $\mu$b & 4.2 $\mu$b & 1.88 mb &  59 mb  \\
 $b\bar{b}$ & 0.32 $\mu$b & 0.2 $\mu$b  & 2.1 $\mu$b &  10 $\mu$b   \\
\hline
\end{tabular}
\caption{Comparison of total cross sections for DPE, two-photon and photon-Pomeron  channels for central charm and bottom production in PbPb  collisions at the LHC.}
\label{tab1}
\end{table}

Finally, we compare the present calculation with similar channels of production. Let us consider first the exclusive photoproduction of heavy quarks in ultra-peripheral collisions in heavy ion collisions \cite{upcs}, which is dominated by photon-Pomeron interactions. The reason is that the final state configuration is similar to the coherent diffraction (both nuclei remain intact and one rapidity gap). It has been recently computed in Ref. \cite{pp1} for proton-proton collisions. It was found $\sigma^{\mathrm{upc}}_{pp}= 0.16-0.53$ $\mu$b for charm and $\sigma^{\mathrm{upc}}_{pp}= 0.32-3$ nb for bottom, where the band correspond to model dependence using distinct saturation models. These values can be compared to Table I and it is verified that the diffractive channel dominates over exclusive photoproduction channel in hadronic collisions. On the other hand, for heavy ion collisions it was found in Ref. \cite{pp2} that $\sigma^{\mathrm{upc}}_{\mathrm{PbPb}}= 59$ mb for charm and $\sigma^{\mathrm{upc}}_{\mathrm{PbPb}}= 10$ $\mu$b for bottom (lower bound cross section relying on saturation physics). Therefore, for heavy ions the photon-Pomeron channel should dominate over the diffractive channel (see Table II). Here, some comments are in order. In general, the parton saturation models underestimate the experimental results for heavy quark photoproduction at DESY-HERA regime. The reason is that they present a smaller effective power on energy when compared to models including QCD evolution in gluon distribution. Therefore, the results coming from saturation physics can be considered a lower bound for the ultra-peripheral cross sections. The enhancement of UPC's compared to single diffractive channel in $AA$ reactions is easily understood from their distinct $A$-dependences: photoproduction grows as $\propto A^4$ ($Z^2$ enhancement from the equivalent photon flux plus an additional enhancement from the photonuclear cross section, $\sigma_{\gamma A}^{\mathrm{exc}}\propto A^{4/3}$), whereas single diffractive cross section gets a factor $A_{\mathrm{eff}}^2\simeq A^{1/3}$. Another channel producing similar final state configuration is the two-photon channel. For instance, it has been calculated in Ref. \cite{PomUPCII} using the color dipole formalism for the photon-photon interaction. The numerical results for these calculations are $\sigma_{\gamma \gamma}^{\mathrm{PbPb}}= 1.81-1.95$ mb for charm and $\sigma^{\mathrm{PbPb}}_{\gamma \gamma}= 2-2.2$ $\mu$b for bottom, using $m_c=1.3$ GeV and $m_b=4.5$ GeV. The lower value correspond to a parton saturation model and the upper one to the BFKL approach. The average value for the 2-photon cross section is presented in Table III. Thus, for heavy ions collisions also the two-photon induced processes are larger than the diffractive channel. For experimental separation of distinct channels, further kinematical cuts have to be imposed. As a summary of distinct production channels, in Table III the cross sections are presented for PbPb collisions.

In summary, we have presented predictions for the inclusive DPE heavy flavor production in heavy ion collisions at the LHC. The cross sections are large enough and can be investigated experimentally. The hard diffraction factorization  was considered further corrected by absorption corrections given by gap survival probability factor. For the Pomeron structure function, the H1 diffractive parton density functions were considered. We investigate the theoretical uncertainty on the multiple interaction corrections for the nuclear case and addressed the diffractive coherent scattering. The main results are the estimations for PbPb collisions, where it has been obtained  $\sigma^{\mathrm{DPE}}_{\mathrm{PbPb}}=32.5\,(0.32) $ $\mu$b  for charm (bottom).  We verified that the diffractive channel dominates over exclusive photoproduction channel in proton-proton case, whereas it is two or more orders of magnitude smaller than the photon-Pomeron contribution in heavy-ion collisions. The enhancement of ultra-peripheral collisions  compared to single diffractive channel  is driven by their distinct $A$-dependences in each process. The two-photon contribution is also important and it is clear that an accurate study using relevant kinematic cuts (rapidity gap separation and transverse momentum spectrum) in the LHC is in order.


\section*{Acknowledgments}

One of us (MVTM) acknowledges the hospitality of \'Ecole Polytechnique, Palaiseau, and the organizers of the workshop {\it Quarkonium 2010: Three days of quarkonium production in $pp$ and $pA$ collisions} (Paliseau, France. July 29-31, 2010) for their invitation, where this work was accomplished. This work was supported by CNPq and FAPERGS, Brazil.


%
%


\end{document}